\documentclass[draftcls,12pt,onecolumn,oneside]{IEEEtran}
\usepackage{mathrsfs}
\usepackage{algorithm}
\usepackage{algorithmic}
\usepackage{url,times,amsmath,color,amssymb,graphicx,epsfig,psfig,cite,geometry,psfrag,subfigure,dsfont,booktabs,amsthm}

\newtheorem{defn}{Definition}
\newtheorem{lem}{Lemma}
\newtheorem{thm}{Theorem}

\begin{document}
%文章题目
\title{Downlink Energy Efficiency of Power Allocation and Wireless Backhaul Bandwidth Allocation in Heterogeneous Small Cell Networks}
%作者
\author{Haijun Zhang,~\IEEEmembership{Senior Member,~IEEE}, Hao Liu, Julian Cheng,~\IEEEmembership{Senior Member,~IEEE}, and Victor C. M. Leung,~\IEEEmembership{Fellow,~IEEE}
%作者简介
\thanks{
Haijun Zhang is  with the Beijing Engineering and Technology Research Center for Convergence Networks and Ubiquitous Services, University of Science and Technology Beijing, Beijing, 100083, China (e-mail: haijunzhang@ieee.org).

Hao Liu and Julian Cheng are with School of Engineering, The University of British Columbia, Kelowna, BC, Canada (e-mail: {hao.liu@alumni.ubc.ca, julian.cheng@ubc.ca}).

Victor C. M. Leung is with the Department of Electrical and Computer Engineering, The University of British Columbia, Vancouver, BC V6T 1Z4 Canada (e-mail: vleung@ece.ubc.ca).
}}
\maketitle

\vspace{-15mm}
%摘要内容
\begin{abstract}
The widespread application of wireless services and dense devices access have triggered huge energy consumption. Because of the environmental and financial considerations, energy-efficient design in wireless networks becomes an inevitable trend. To the best of the authors' knowledge, energy-efficient orthogonal frequency division multiple access heterogeneous small cell optimization comprehensively considering energy efficiency maximization, power allocation, wireless backhaul bandwidth allocation, and user Quality of Service is a novel approach and research direction, and it has not been investigated. In this paper, we study the energy-efficient power allocation and wireless backhaul bandwidth allocation in orthogonal frequency division multiple access heterogeneous small cell networks. Different from the existing resource allocation schemes that maximize the throughput, the studied scheme maximizes energy efficiency by allocating both transmit power of each small cell base station to users and bandwidth for backhauling, according to the channel state information and the circuit power consumption. The problem is first formulated as a non-convex nonlinear programming problem and then it is decomposed into two convex subproblems. A near optimal iterative resource allocation algorithm is designed to solve the resource allocation problem. A suboptimal low-complexity approach is also developed by exploring the inherent structure and property of the energy-efficient design. Simulation results demonstrate the effectiveness of the proposed algorithms by comparing with the existing schemes.
\end{abstract}
\vspace{-5mm}
%关键字
\begin{keywords}
Bandwidth allocation, energy efficiency, heterogeneous network, power allocation, small cell, wireless backhaul.
\end{keywords}

%introduction
\section{Introduction}

Wireless communication networks have experienced tremendous growth in the past a few decades. It is shown that higher capacity wireless links are expected to meet the increasing quality of service (QoS) demands of multimedia applications, and these high data rate links also result in increasing device power consumption. The next generation communication systems need to provide higher data rate with limited power and bandwidth due to the rapidly increasing demands for multimedia services. Designing energy-efficient wireless communication system becomes an emerging trend, due to rapidly increasing system energy costs and rising requirements of communication capacity \cite{Energy-Efficient-Non-cooperative14, Green-Communications12, Pricing-based-multiresource-allocation-in-OFDMA14}. According to \cite{Energy-Efficiency-Enhancements04} and \cite{Fundamental-trade-offs11}, the radio access part is a major energy consumer in conventional wireless cellular networks, and it accounts for up to more than 70 percent of the total energy consumption. Therefore, increasing the energy efficiency of typical wireless networks is important to overcome the challenges raised by the rising demands of energy consumption and communication throughput.

To offload the overloaded traffics in macrocells and enhance the capacity and energy efficiency of the wireless networks, one proposed method is to shorten the distance between the base stations (BSs) and the user equipments. Small cells (e.g., picocells, femtocells and relay nodes) have been used to improve system capacity in hotspots for relieving the burden on overloaded macrocells, which is considered as a promising technique to provide an effective solution for the challenges in current macrocells \cite{Coexistence-of-Wi-Fi15, Power-minimization-based-resource-allocation14}. Therefore, there is no doubt that small cell has been paid much attention in recent years from academia and industry because it can help the system spatial reuse spectrum with low power consumption and improve the system coverage with low infrastructure cost deployment \cite{Cooperative-Interference-Mitigation15}. Heterogeneous small cell networks, where small cells are overlaid within a macrocell to improve coverage and increase system capacity beyond the initial deployment of macrocells, have been regarded as a promising approach to meet the increasing data traffic demand and reduce energy consumption. Although highly promising, many important problems related to heterogeneous small cell networks such as interference mitigation, resource allocation, and QoS provisioning \cite{A-Cooperative-Bargaining15, WenChiEnergy2016, Interference-Limited-Resource-Optimization15, Resource-Allocation-in-Spectrum-Sharing-OFDMA14} should be addressed to fully reap the potential gains.

Resource allocation, such as power allocation and bandwidth allocation, has been widely used to maximize the energy efficiency under power limitation and QoS in heterogeneous small cell networks. Power allocation for energy efficiency has been widely studied in the literature. The distributed power control game was studied in \cite{A-Repeated-Game-Formulation10} to maximize the energy efficiency of transmission for secondary users in cognitive radio networks and an optimal power control problem was formulated as a repeated game. In \cite{EE2017}, based on the hardcore point process (HCPP), the authors investigated the maximum achievable energy efficiency of the considered multiuser multiantenna HCPP random cellular networks with the aforementioned minimum distance constraint for adjacent BSs. Different from the authors in \cite{EE2017}, who took the minimum distance in adjacent BSs into consideration to maximize the energy efficient, we propose a suboptimal low-complexity approach of energy-efficient backhaul bandwidth allocation by optimizing the fraction of bandwidth allocated for wireless backhauling at all small cell BSs within a macrocell range. The authors in \cite{A-game-theoretic-approach11} studied energy-efficient power control and receiver design in cognitive radio networks, and a non-cooperative power control game for maximum energy efficiency of secondary users was considered with a fairness constraint and interference threshold. The authors of \cite{Energy-efficient-resource-allocation-for12} formulated the energy-efficient spectrum sharing and power allocation in heterogeneous cognitive radio networks with femtocells as a Stackelberg game and they proposed a gradient based iteration algorithm to obtain the Stackelberg equilibrium solution to the energy-efficient resource allocation problem. Some works also have been done to consider bandwidth allocation for energy efficiency. In \cite{Pricing-and-bandwidth-allocation11}, the authors studied the joint service pricing and bandwidth allocation for energy and cost efficiency at the operator level in a multi-tier network where an operator deploys heterogeneous small cell networks, and they formulated the problem as a Stackelberg game. The problem of joint link selection, power and bandwidth allocation for energy efficiency maximization for Multi-Homing networks was investigated in \cite{Energy-Efficient-Bandwidth15}. A new energy-efficient scheme was presented in \cite{Energy-Efficient-Bandwidth-Allocation06} to statistically meet the demands for QoS during the bandwidth allocation for wireless networks.

In this paper, we define that wireless backhaul as the connection between macro BS and small cell BSs, and it is necessary to jointly consider the design of the radio access and backhaul network. Several related works considered the backhaul to improve energy efficiency in wireless networks. The authors of \cite{Energy-Efficient-Resource-Allocation-in-Multi-Cell12} studied energy efficiency of resource allocation in multi-cell orthogonal frequency division multiple access (OFDMA) downlink networks where the limited backhaul capacity, the circuit power consumption and the minimum required data rate are considered. The resource allocation problem for energy-efficient communication with limited backhaul capacity is formulated as an optimization problem. In \cite{Energy-efficiency-of-small-cell-backhaul-networks15}, an energy efficiency model of small cell backhaul networks with Gauss--Markov mobile models has been proposed. In \cite{Energy-Efficient-Forward-and-Backhaul15}, the authors maximized system energy efficiency in OFDMA small cell networks by optimizing backhaul data rate and emission power, and they proposed a joint forward and backhaul link optimization scheme by taking both the power consumption of forward links and the backhaul links into consideration.

To the best of the authors' knowledge, energy efficiency of power allocation and wireless backhaul bandwidth allocation in heterogeneous small cell network has not been investigated. In this work, we study the power allocation and bandwidth allocation problem in a heterogeneous small cell network where the small cells use wireless backhauling to maximize energy efficiency of all small cell users. Similar to the paper in \cite{Energy-efficient10}, we also use Gradient Assisted Binary Search (GABS) Algorithm to solve the energy-efficient power allocation problem. Reference \cite{HaoLiuICC2016} is a conference version of this paper. Different from the conference version, we provide the detailed proof for the theorem, complexity analysis for the proposed algorithms and more simulation results in this paper. The key contributions of our work can be summarized as follows.

\begin{itemize}
  \item \emph{Design of an energy-efficient OFDMA heterogeneous small cell optimization:} This is a novel approach by considering energy efficiency maximization, power allocation, wireless backhaul bandwidth allocation, and user QoS into the design of OFDMA heterogeneous small cell optimization.
%  \item \emph{Formulation of energy efficiency of wireless backhaul bandwidth allocation and power allocation algorithm with respect to multiple constraints:}
      We formulate the energy-efficient wireless backhaul bandwidth allocation and power allocation problem in a heterogeneous small cell as a nonlinear programming problem, where maximum transmit power constraints of each small cell BS to each small cell user, the downlink data rate constraint of small cell BSs and the minimum data rate between each small cell BS and its corresponding user are considered to provide reliable and low energy consumed downlink transmission for small cell users. The non-convex optimization problem is then decomposed into two convex subproblems, and an algorithm is proposed for wireless backhual bandwidth allocation and power allocation.
  \item \emph{Support of the small cell backhauling in the context of designing power allocation schemes for heterogeneous small cell networks:} We study the wireless backhaul bandwidth allocation at the small cell BS, which means a fraction of bandwidth is scheduled for backhauling and the other is assigned for communication with corresponding users. We formulate the bandwidth allocation problem as a convex problem and obtain the optimum solution.
  \item \emph{Design of suboptimal low-complexity algorithm by decomposing the power allocation and bandwidth allocation:} The energy-efficient wireless backhaul bandwidth allocation and power allocation problem are decomposed and are optimized separately. Correspondingly, a suboptimal low-complexity algorithm is proposed. The effectiveness of the proposed suboptimal algorithm is demonstrated by simulations.
\end{itemize}

The rest of this paper is organized as follows. Section II describes the system model. In Section III, the energy-efficient resource allocation and backhauling are presented, and in Section IV, optimization algorithms are proposed. Simulation results are discussed in Section V. Finally, Section VI concludes the paper.

%system model
\section{System Model}

In this section, we formulate the problem of downlink energy efficiency of power allocation and unified bandwidth allocation for wireless backhauling in heterogeneous small cell networks.

We consider a heterogeneous small cell network as shown in Fig. 1 with a single macro BS, $J$ small cells deployed within the macrocell range and $K$ users randomly located in each small cell.

The small cells share the same spectrum with macrocell. In this work, the unified wireless backhaul bandwidth allocation is investigated. The unified bandwidth allocation factor $\beta  \in [0,1]$, which is the fraction of bandwidth allocated for wireless backhauling at all small cell BSs within a macrocell range. For simplicity, all small cells are assumed to have the same bandwidth allocation factor. We assume that the multiple antenna technology is used in the macro BS and each small cell corresponds to a beamforming group, so the interference for wireless backhaul between different small cells can be neglected. The antenna array size at macro BS is $N$, which is much greater than the beamforming group size $B$ and the number of small cells, i.e., $N \gg B$ and $N \gg J$. In this work, we also assume that $B \ge J$. Each small cell BS is equipped with single antenna. OFDMA technology is used in each small cell to support the communication between BS and users.

Let ${P_0}$ be the equal transmit power of the macro BS transmit antenna targeted at corresponding small cell and ${\sigma ^2}$ is the additive white Gaussian noise (AWGN) power. Then the received signal-to-noise ratio (SNR) in the wireless backhaul downlink of small cell $j$ is given by

  \begin{equation}
  {\gamma _j} = \frac{{{P_0}{G_j}}}{{{\sigma ^{\rm{2}}}}}.
  \end{equation}

  Let ${g_{j,k}}$ be the channel power gain between the $j$th small cell BS and its corresponding $k$th user,  where $j \in \{ 1,2,...,J\}$, $k \in \{ 1,2,...,K\}$. Let ${p_{j,k}}$ denote the transmit power from the $j$th small cell BS to its corresponding $k$th user, and let ${\bf{P}} = {[{p_{j,k}}]_{J \times K}}$ denote the power allocation matrix.

We assume that different users in each small cell use different subchannels and co-channel interference between small cells as part of the thermal noise because of the severe wall penetration loss and low power of small cell BSs \cite{Resource-Allocation-in-Spectrum-Sharing-OFDMA14}. The received signal-to-interference-plus-noise ratio (SINR) of small cell user $k$ associated with small cell $j$ is given by
\begin{equation}
{\gamma _{j,k}} = \frac{{{p_{j,k}}{g_{j,k}}}}{{{\sigma ^{\rm{2}}}{\rm{ + }}{I_{j,k}}}}
\end{equation}
where ${I_{j,k}}$ is the interference introduced by macro BS, ${I_{j,k}} = {P_0}{G_{j,k}}$, where ${G_{j,k}}$ is the channel power gain between macro BS and the $k$th user in the $j$th small cell. The achievable data transmission rate between the $j$th small cell BS and its corresponding $k$th user is determined by
\begin{equation}
{r_{j,k}} = \left( {\frac{{1 - \beta }}{K}} \right){\log _2}\left( {1 + {\gamma _{j,k}}} \right).
\end{equation}
Therefore, we have the relation between ${r_{j,k}}$ and ${p_{j,k}}$

\begin{equation}
\label{r&p relation}
\begin{aligned}
&&& {p_{j,k}} = ({2^{\frac{{K{r_{j,k}}}}{{1 - \beta }}}} - 1)\frac{{{\sigma ^{\rm{2}}}{\rm{ + }}{I_{j,k}}}}{{{g_{j,k}}}} \\
&&& {r_{j,k}} = \left( {\frac{{1 - \beta }}{K}} \right){\log _2}\left( {1 + \frac{{{p_{j,k}}{g_{j,k}}}}{{{\sigma ^{\rm{2}}}{\rm{ + }}{I_{j,k}}}}} \right).\\
\end{aligned}
\end{equation}

Besides the transmit power during the transmission, circuit energy consumption is also incurred by device electronics in small cell BSs \cite{Energy-constrained-modulation-optimization05, Energy-efficient01}. Circuit power represents the additional device power consumption of devices during transmissions \cite{Energy-efficiency04}, such as digital-to-analog converters, mixers and filters, and this portion of energy consumption is independent of the transmission state. If we denote the circuit power as ${P_C}$ , the overall power assumption of the $j$th small cell BS to the $k$th user is ${P_C} + {p_{j,k}}$.

For energy-efficient communication, it is desirable to send the maximum amount of data with a given amount of energy for small cell BSs. Hence, given any amount of energy $\Delta e$ consumed in a duration $\Delta t$ in each small cell BS to each user, $\Delta e = \Delta t({P_C} + {p_{j,k}})$, the small cell BSs desire to send a maximum amount of data by choosing the power allocation vector and backhaul bandwidth to maximize
\begin{equation}
\sum\limits_{j = 1}^J {\sum\limits_{k = 1}^K {\frac{{{r_{j,k}}(\beta ,{p_{j,k}})\Delta t}}{{\Delta e}}} }
\end{equation}
which is equivalent to maximizing
\begin{equation}
U(\beta ,{\bf{P}}) = \sum\limits_{j = 1}^J {\sum\limits_{k = 1}^K {{U_{j,k}}(\beta ,{p_{j,k}})} }
\end{equation}
where
\begin{equation}
{U_{j,k}}(\beta ,{p_{j,k}}) = \frac{{{r_{j,k}}(\beta ,{p_{j,k}})}}{{{P_C} + {p_{j,k}}}}.
\end{equation}
In (6), $U(\beta ,{\bf{P}})$ is called energy efficiency for all small cells; ${U_{j,k}}(\beta ,{p_{j,k}})$ is the energy efficiency of the $k$th user of the $j$th small cell. The unit of the energy efficiency is bits per Hertz per Joule, which has been frequently used in the literature for energy-efficient communications \cite{Spectral-efficiency02, An-energy-efficient06, Power-limited-channels88, Power-control00, Pricing-and-power-control04}.

When the downlink channel state information is estimated by the small cell BSs, the resource allocation is performed by a small cell BS under the following constraints.
\begin{itemize}
  \item \emph{Transmit power constraint of each small cell BS to each user:}
    \begin{equation}
    0 \le {p_{j,k}} \le {P_{\max }},\forall j,k
    \end{equation}
    where ${P_{\max }}$ denotes the maximal transmit power of each small cell BS to each user.
  \item \emph{The downlink data rate constraint of each small cell BS:}
    The throughput of the  small cell is given by
    \begin{equation}
    {R_j} = \sum\limits_{k = 1}^K {{r_{j,k}}} .
    \end{equation}
    Due to the inter-user interference within the overlapping areas of macrocell and small cell beamforming group, we use typical zero-forcing beamforming technique with equal-power allocation for each user transmission link to eliminate the interference significantly \cite{Wangning2016, ITA2014}, the capacity of the wireless backhaul downlink for small cell $j$ is
    \begin{equation}
    {C_j} = \beta {\log _2}\left( {1 + \frac{{N{\rm{ - }}B{\rm{ + 1}}}}{B}{\gamma _j}} \right).
    \end{equation}
    The downlink wireless backhaul constraint requires
    \begin{equation}
    {R_j} \le {C_j}
    \end{equation}
    such that the downlink traffic of the $j$th small cell can be accommodated by its wireless backhaul.
  \item \emph{Heterogeneous QoS guarantee:}
    The QoS requirement ${R_t}$ should be guaranteed for each user in each small cell to maintain the performance of the communication system
    \begin{equation}
    {r_{j,k}} \ge {R_t}.
    \end{equation}
\end{itemize}

Our target is to maximize the energy efficiency of power allocation and unified bandwidth allocation for wireless backhauling in heterogeneous small cell networks under power constraint and data rate requirements. Thus, the corresponding problem for the downlink can be formulated as the following nonlinear programming problem
\begin{equation}
\label{originalobj}
\mathop {\max }\limits_{\beta ,{\bf{P}}} U(\beta ,{\bf{P}}) = \mathop {\max }\limits_{\beta ,{p_{j,k}}} \sum\limits_{j = 1}^J {\sum\limits_{k = 1}^K {{U_{j,k}}(\beta ,{p_{j,k}})} }
\end{equation}
\begin{equation}
\label{originalconstraints}
\begin{aligned}
& \text{s.t.}
& & {C1:0 \le {p_{j,k}} \le {P_{\max }}} \\
&&& {C2:{R_j} \le {C_j}} \\
&&& {C3:{r_{j,k}} \ge {R_t}}\\
&&& {C4:0 \le \beta  \le 1}.\\
\end{aligned}
\end{equation}

%Energy-efficient Resource Allocation and Backhauling
\section{Energy-efficient Resource Allocation and Backhauling}

Since the optimization problem formulated in (\ref{originalobj}) and (\ref{originalconstraints}) is non-convex and we notice that the continuous variables $\beta$ and ${p_{j,k}}$ are separable in (\ref{originalobj}). Therefore, we consider a decomposition approach to solve the energy-efficient resource allocation problem. We decompose the non-convex optimization problem into two convex subproblems: one for energy-efficient power allocation and one for energy-efficient wireless backhaul bandwidth allocation. Then, we solve the subproblems of energy-efficient power allocation and energy-efficient backhaul bandwidth allocation individually.

\subsection{Energy-Efficient Power Allocation}

The concept of quasiconcavity will be used in our discussion and is defined in \cite{Topics-in-Microeconomics99}.
%definition
\begin{defn}
A function $f$, which maps from a convex set of real n-dimensional vectors, $D$, to a real number, is called strictly quasiconcave if for any ${x_1},{x_2} \in D$ and ${x_1} \ne {x_2}$,
\begin{equation}
f(\lambda {x_1} + (1 - \lambda ){x_2}) > \min \{ f({x_1}),f({x_2})\}
\end{equation}
for any $0 < \lambda  < 1$.
\end{defn}

Given a value $\beta $ for unified wireless backhaul bandwidth allocation, the optimization algorithm begins with the power allocation subproblem P1.1 that is formulated as
\begin{equation}
\label{P1.1objectivefunc}
{\rm{P1}}{\rm{.1}}:\mathop {\max }\limits_{\bf{P}} U({\bf{P}}) = \mathop {\max }\limits_{{p_{j,k}}} \sum\limits_{j = 1}^J {\sum\limits_{k = 1}^K {{U_{j,k}}({p_{j,k}})} }
\end{equation}
\begin{equation}
\label{P1.1constraints}
\begin{aligned}
& \text{s.t.}
& & {C1:0 \le {p_{j,k}} \le {P_{\max }}} \\
&&& {C2:{R_j} \le {C_j}} \\
&&& {C3:{r_{j,k}}({p_{j,k}}) \ge {R_t}}\\
\end{aligned}
\end{equation}
where
\begin{equation}
{r_{j,k}}({p_{j,k}}) = \left( {\frac{{1 - \beta }}{K}} \right){\log _2}\left( {1 + \frac{{{p_{j,k}}{g_{j,k}}}}{{{\sigma ^{\rm{2}}}{\rm{ + }}{I_{j,k}}}}} \right)
\end{equation}
is strictly concave and monotonically increasing in ${p_{j,k}}$ with ${r_{j,k}}(0) = 0$, when ${p_{j,k}} = 0$.

The optimal energy-efficient power allocation achieves the maximum energy efficiency, i.e.
\begin{equation}
{{\bf{P}}^ * } = \mathop {\arg \max }\limits_{\bf{P}} U({\bf{P}}).
\end{equation}
It is proved in Appendix A that $U({\bf{P}})$ has the following properties.

%lemma
\begin{lem}
If ${r_{j,k}}({p_{j,k}})$ is strictly concave in ${p_{j,k}}$, ${U_{j,k}}({p_{j,k}}) \in U({\bf{P}})$ is strictly quasiconcave. Furthermore, ${U_{j,k}}({p_{j,k}})$ is first strictly increasing and then strictly decreasing in any ${p_{j,k}}$, i.e. the local maximum of $U({\bf{P}})$ for each ${p_{j,k}}$ exists at a positive finite value.
\end{lem}

For strictly quasiconcave functions, if a local maximum exists, it is also globally optimal \cite{Topics-in-Microeconomics99}. Hence, a unique globally optimal transmission rate vector always exists and its characteristics are summarized in Theorem 1 according to the proofs in Appendix A.

%theorem
\begin{thm}
If ${r_{j,k}}({p_{j,k}})$ is strictly concave, there exists a unique globally optimal transmission power vector ${{\bf{P}}^ * } = \{ {p^ * }_{j,k};(j,k) \in J \times K\} $ for ${{\bf{P}}^ * }{\rm{ = }}\mathop {\arg \max }\limits_{\bf{P}} U({\bf{P}})$, for each element in ${{\bf{P}}^*}$, $p_{j,k}^*{\rm{ = }}\mathop {\arg \max }\limits_{{p_{j,k}}} {U_{j,k}}({p_{j,k}})$ where $p_{j,k}^*$ is given by
\[\begin{array}{*{20}{l}}
{{{\left. {\frac{{\partial {U_{j,k}}({p_{j,k}})}}{{\partial {p_{j,k}}}}} \right|}_{{p_{j,k}} = p_{j,k}^*}} = 0,\;f({p_{j,k}}) = 0,}\\
{i.e.,\;{U_{j,k}}(p_{j,k}^*) = \frac{{{r_{j,k}}(p_{j,k}^*)}}{{{P_C} + p_{j,k}^*}} = {{\left. {\frac{{\partial {r_{j,k}}({p_{j,k}})}}{{\partial {p_{j,k}}}}} \right|}_{{p_{j,k}} = p_{j,k}^*}}.}
\end{array}\]
\end{thm}

In order to solve the problem P1.1 for power allocation, we rewrite the objective function in (\ref{P1.1objectivefunc}) as
\begin{equation}
\label{rewriteobjfunc}
\mathop {\max }\limits_{{p_{j,k}}} {U_{j,k}}({p_{j,k}}) = \mathop {\max }\limits_{{p_{j,k}}} \frac{{{r_{j,k}}({p_{j,k}})}}{{{P_C} + {p_{j,k}}}}.
\end{equation}
If each small cell user could reach the maximum energy efficiency, all small cell users could reach the maximum energy efficiency. The total data rate in each small cell could not exceed the capacity of the wireless backhaul downlink for small cell $j$, ${R_j} \le {C_j}$, we can approximate that the data rate for each user to be less than $\frac{{{C_j}}}{K}$, ${r_{j,k}}({p_{j,k}}) \le \frac{{{C_j}}}{K}$, and the maximum of power for user $k$ in small cell $j$ is $\frac{{{P_S}}}{K}$.
     Thus, P1.1 is equivalent to
     \begin{equation}
     {\rm{P1}}{\rm{.2}}:\mathop {\max }\limits_{{p_{j,k}}} {U_{j,k}}({p_{j,k}})
     \end{equation}
     \begin{equation}
     \begin{aligned}
     \label{P1.3constraints}
     & \text{s.t.}
     & & {C1:0 \le {p_{j,k}} \le \frac{{{P_S}}}{K}} \\
     &&& {C2:{r_{j,k}}({p_{j,k}}) \le \frac{{{C_j}}}{K}} \\
     &&& {C3:{r_{j,k}}({p_{j,k}}) \ge {R_t}}.\\
     \end{aligned}
     \end{equation}

We can rewrite \emph{C2} in (\ref{P1.3constraints}) according to (\ref{r&p relation}) as
%\begin{equation}
%\left( {\frac{{1 - \beta }}{K}} \right){\log _2}\left( {1 + \frac{{{p_{j,k}}{g_{j,k}}}}{{{N_0}{\rm{ + }}{I_{j,k}}}}} \right) \le \left( {\frac{\beta }{K}} \right){\log _2}\left( {1 + \frac{{N{\rm{ - }}B{\rm{ + 1}}}}{B}\frac{{{P_0}{G_j}}}{{{N_0}}}} \right)
%\end{equation}
%\begin{equation}
%{\log _2}\left( {1 + \frac{{{p_{j,k}}{g_{j,k}}}}{{{N_0}{\rm{ + }}{I_{j,k}}}}} \right) \le \left( {\frac{\beta }{{1 - \beta }}} \right){\log _2}\left( {1 + \frac{{N{\rm{ - }}B{\rm{ + 1}}}}{B}\frac{{{P_0}{G_j}}}{{{N_0}}}} \right)
%\end{equation}
%\begin{equation}
%\frac{{{p_{j,k}}{g_{j,k}}}}{{{N_0}{\rm{ + }}{I_{j,k}}}} \le {{\rm{2}}^{\left( {\frac{\beta }{{1 - \beta }}} \right){{\log }_2}\left( {1 + \frac{{N{\rm{ - }}B{\rm{ + 1}}}}{B}\frac{{{P_0}{G_j}}}{{{N_0}}}} \right)}}{\rm{ - 1}}
%\end{equation}
\begin{equation}
{p_{j,k}} \le \left( {\frac{{{\sigma ^{\rm{2}}}{\rm{ + }}{I_{j,k}}}}{{{g_{j,k}}}}} \right)\left( {{{\rm{2}}^{\left( {\frac{\beta }{{1 - \beta }}} \right){{\log }_2}\left( {1 + \frac{{N{\rm{ - }}B{\rm{ + 1}}}}{B}\frac{{{P_0}{G_j}}}{{{\sigma ^{\rm{2}}}}}} \right)}}{\rm{ - 1}}} \right).
\end{equation}
We can rewrite \emph{C3} in (\ref{P1.3constraints}) according to (\ref{r&p relation}) as
%\begin{equation}
%\left( {\frac{{1 - \beta }}{K}} \right){\log _2}\left( {1 + \frac{{{p_{j,k}}{g_{j,k}}}}{{{N_0}{\rm{ + }}{I_{j,k}}}}} \right) \ge {R_t}
%\end{equation}
\begin{equation}
{p_{j,k}} \ge \left( {\frac{{{\sigma ^{\rm{2}}}{\rm{ + }}{I_{j,k}}}}{{{g_{j,k}}}}} \right)\left( {{{\rm{2}}^{\frac{{K{R_t}}}{{1 - \beta }}}}{\rm{ - 1}}} \right).
\end{equation}
Therefore, we have
\begin{equation}
{L_{j,k}} \le {p_{j,k}} \le {H_{j,k}}
\end{equation}
where
\begin{equation}
{L_{j,k}} = \left( {\frac{{{\sigma ^{\rm{2}}}{\rm{ + }}{I_{j,k}}}}{{{g_{j,k}}}}} \right)\left( {{{\rm{2}}^{\frac{{K{R_t}}}{{1 - \beta }}}}{\rm{ - 1}}} \right)
\end{equation}
\begin{equation}
{H_{j,k}} = \min \left\{ {\left( {\frac{{{\sigma ^{\rm{2}}}{\rm{ + }}{I_{j,k}}}}{{{g_{j,k}}}}} \right)\left( {{{\rm{2}}^{\left( {\frac{\beta }{{1 - \beta }}} \right){{\log }_2}\left( {1 + \frac{{N{\rm{ - }}B{\rm{ + 1}}}}{B}\frac{{{P_0}{G_j}}}{{{\sigma ^{\rm{2}}}}}} \right)}}{\rm{ - 1}}} \right),{P_{\max }}} \right\}
\end{equation}
only if the following inequality is satisfied
\begin{equation}
{L_{j,k}} \le {H_{j,k}}.
\end{equation}

The energy-efficient power allocation is given by
\begin{equation}
{{\hat p}^*}_{j,k} = \mathop {\arg \max }\limits_{{p_{j,k}}} \frac{{{r_{j,k}}({p_{j,k}})}}{{{P_C} + {p_{j,k}}}}
\end{equation}
subject to
\begin{equation}
{L_{j,k}} \le {p_{j,k}} \le {H_{j,k}}.
\end{equation}

We can solve (\ref{rewriteobjfunc}) by using Theorem 1 to find the optimal power allocation solution. We can also use the low-complexity iterative algorithms based on the GABS algorithm proposed in \cite{Energy-efficient10} to realize the energy-efficient power allocation for the $k$th user in the $j$th small cell BS as follows.

{\renewcommand\baselinestretch{1.0}\small
\begin{algorithm}[!h]\vspace{-1pt}
\renewcommand{\thealgorithm}{}
\caption{GABS Algorithm}
\begin{algorithmic}[1]{\small}
\STATE  \textbf{Initialization:} Each small cell BS allocates the same transmit power to each user, ${p_{j,k}} > 0$.
\STATE  Then do $p_{j,k}^{(1)} = {p_{j,k}}$, ${h_1} \leftarrow \frac{{\partial {U_{j,k}}({p_{j,k}})}}{{\partial {p_{j,k}}}}\left| {_{{p_{j,k}} = p_{j,k}^{(1)}}} \right.$ and $c > 1$ (let $c = 2$).
\IF     {${h_1} < 0$}
\REPEAT
\STATE  $p_{j,k}^{({\rm{2}})} \leftarrow p_{j,k}^{(1)}$, $p_{j,k}^{(1)} \leftarrow \frac{{p_{j,k}^{(1)}}}{c}$, and ${h_1} \leftarrow \frac{{\partial {U_{j,k}}({p_{j,k}})}}{{\partial {p_{j,k}}}}\left| {_{{p_{j,k}} = p_{j,k}^{(1)}}} \right.$
\UNTIL  {${h_1} \ge 0$}
\ELSE
\STATE  $p_{j,k}^{({\rm{2}})} \leftarrow p_{j,k}^{(1)} \times c$ and ${h_2} \leftarrow \frac{{\partial {U_{j,k}}({p_{j,k}})}}{{\partial {p_{j,k}}}}\left| {_{{p_{j,k}} = p_{j,k}^{({\rm{2}})}}} \right.$
\REPEAT
\STATE  $p_{j,k}^{(1)} \leftarrow p_{j,k}^{({\rm{2}})}$, $p_{j,k}^{({\rm{2}})} \leftarrow p_{j,k}^{({\rm{2}})} \times c$ and ${h_2} \leftarrow \frac{{\partial {U_j}({p_{j,k}})}}{{\partial {p_{j,k}}}}\left| {_{{p_{j,k}} = p_{j,k}^{({\rm{2}})}}} \right.$
\UNTIL  {${h_2} \le 0$}
\ENDIF
\WHILE{no convergence}
\STATE  ${{\hat p}^*}_{j,k} \leftarrow \frac{{p_{j,k}^{(1)} + p_{j,k}^{({\rm{2}})}}}{2}$, $h' \leftarrow \frac{{\partial {U_{j,k}}({p_{j,k}})}}{{\partial {p_{j,k}}}}\left| {_{{p_{j,k}} = {{\hat p}^*}_{j,k}}} \right.$
\IF     {$h' > 0$}
\STATE  $p_{j,k}^{(1)} = {{\hat p}^*}_{j,k}$
\ELSE
\STATE  $p_{j,k}^{({\rm{2}})} = {{\hat p}^*}_{j,k}$
\ENDIF
\ENDWHILE
\STATE \textbf{Output} ${\hat p^ * }_{j,k}$.
\end{algorithmic}
\end{algorithm}
\par}

If the output ${\hat p^ * }_{j,k}$ satisfies the power constraint, i.e. $\hat p_{j,k}^*{\rm{ = }}p_{j,k}^*$; otherwise, we can obtain the maximum ${U_{j,k}}({p_{j,k}})$ by
\begin{equation}
p_{j,k}^* = {L_{j,k}}
\end{equation}
if ${{\hat p}^*}_{j,k} < {L_{j,k}}$, or we can get the maximum ${U_{j,k}}({p_{j,k}})$ by
\begin{equation}
p_{j,k}^* = {H_{j,k}}
\end{equation}
if ${{\hat p}^*}_{j,k} > {H_{j,k}}$, since ${U_{j,k}}({p_{j,k}})$ is first strictly increasing and then strictly decreasing in any positive finite ${p_{j,k}}$.

\subsection{Energy-Efficient Wireless Backhaul Bandwidth Allocation}

Once the optimal solution ${{\bf{P}}^*} = \{ p_{j,k}^*;(j,k) \in J \times K\}$ is obtained for the convex subproblem P1.2 parameterized by $\beta $, it can be used in the following subproblem P1.3 for the unified wireless backhaul bandwidth allocation
\begin{equation}
\label{P1.2obj}
{\rm{P1}}.3:\mathop {\max }\limits_\beta  U(\beta ,{{\bf{P}}^*}) = \mathop {\max }\limits_\beta  \sum\limits_{j = 1}^J {\sum\limits_{k = 1}^K {{U_{j,k}}(\beta ,p_{j,k}^*)} }
\end{equation}

\begin{equation}
\label{P1.2constraints}
\begin{aligned}
& \text{s.t.}
& & {C1:0 \le \beta  \le 1} \\
&&& {C2:{R_j}(\beta ,{{\bf{P}}^ * }) \le {C_j}(\beta ,{{\bf{P}}^ * })} \\
&&& {C3:{r_{j,k}}(\beta ,p_{j,k}^*) \ge {R_t}}\\
\end{aligned}
\end{equation}
where ${R_j}(\beta ,{{\bf{P}}^ * })$ is the function value of ${R_j}$ evaluated at ${{\bf{P}}^ * }$, ${C_j}(\beta ,{{\bf{P}}^ * })$ is the function value of ${C_j}$ evaluated at ${{\bf{P}}^ * }$. In order to obtain the solution to the original problem in (\ref{originalobj}) and (\ref{originalconstraints}), the two subproblems P1.2 and P1.3 are solved iteratively until convergence.

Maximizing the objective function of P1.3 with respect to $\beta $ is equivalent to maximizing $\left( {1 - \beta } \right)$ only, since (\ref{P1.2obj}) is a monotonically decreasing function of $\beta $. Problem P1.3 reduces to a feasibility problem whose solution is the smallest feasible value of $\beta $ given constraints (\ref{P1.2constraints}).

According to \emph{C2}, ${R_j}(\beta ,{{\bf{P}}^ * }) \le {C_j}(\beta ,{{\bf{P}}^ * })$, we have
%\begin{equation}
%\left( {1 - \beta } \right)\sum\limits_{k = 1}^K {\left( {\frac{1}{K}} \right){{\log }_2}\left( {1 + \frac{{p_{j,k}^*{g_{j,k}}}}{{{N_0} + {I_{j,k}}}}} \right)}  \le \beta {\log _2}\left( {1 + \frac{{N{\rm{ - }}B{\rm{ + 1}}}}{B}\frac{{{P_0}{G_j}}}{{{N_0}}}} \right)
%\end{equation}
%\begin{equation}
%\frac{{1 - \beta }}{\beta } \le \frac{{K{{\log }_2}\left( {1 + \frac{{N{\rm{ - }}B{\rm{ + 1}}}}{B}\frac{{{P_0}{G_j}}}{{{N_0}}}} \right)}}{{\sum\limits_{k = 1}^K {{{\log }_2}\left( {1 + \frac{{p_{j,k}^*{g_{j,k}}}}{{{N_0} + {I_{j,k}}}}} \right)} }}
%\end{equation}
\begin{equation}
\beta  \ge \frac{{\sum\limits_{k = 1}^K {{{\log }_2}\left( {1 + \frac{{p_{j,k}^*{g_{j,k}}}}{{{\sigma ^{\rm{2}}} + {I_{j,k}}}}} \right)} }}{{K{{\log }_2}\left( {1 + \frac{{N{\rm{ - }}B{\rm{ + 1}}}}{B}\frac{{{P_0}{G_j}}}{{{\sigma ^{\rm{2}}}}}} \right) + \sum\limits_{k = 1}^K {{{\log }_2}\left( {1 + \frac{{p_{j,k}^*{g_{j,k}}}}{{{\sigma ^{\rm{2}}} + {I_{j,k}}}}} \right)} }}.
\end{equation}

According to \emph{C3}, ${r_{j,k}}(\beta ,p_{j,k}^*) \ge {R_t}$, we have
%\begin{equation}
%\left( {\frac{{1 - \beta }}{K}} \right){\log _2}\left( {1 + \frac{{p_{j,k}^*{g_{j,k}}}}{{{N_0} + {I_{j,k}}}}} \right) \ge {R_t}
%\end{equation}
\begin{equation}
\beta  \le 1 - \frac{{K{R_t}}}{{{{\log }_2}\left( {1 + \frac{{p_{j,k}^*{g_{j,k}}}}{{{\sigma ^{\rm{2}}} + {I_{j,k}}}}} \right)}}.
\end{equation}

So we have
\begin{equation}
\label{computebeta}
\beta  = \max \left\{ {{\phi _j}} \right\}
\end{equation}
where
\begin{equation}
\label{phi _j}
{\phi _j} = \frac{{\sum\limits_{k = 1}^K {{{\log }_2}\left( {1 + \frac{{p_{j,k}^*{g_{j,k}}}}{{{\sigma ^{\rm{2}}} + {I_{j,k}}}}} \right)} }}{{K{{\log }_2}\left( {1 + \frac{{N{\rm{ - }}B{\rm{ + 1}}}}{B}\frac{{{P_0}{G_j}}}{{{\sigma ^{\rm{2}}}}}} \right) + \sum\limits_{k = 1}^K {{{\log }_2}\left( {1 + \frac{{p_{j,k}^*{g_{j,k}}}}{{{\sigma ^{\rm{2}}} + {I_{j,k}}}}} \right)} }}
\end{equation}
only if ${R_t}$ satisfies the following condition
%\begin{equation}
%1 - \frac{{K{R_t}}}{{{{\log }_2}\left( {1 + \frac{{p_{j,k}^*{g_{j,k}}}}{{{N_0} + {I_{j,k}}}}} \right)}} \ge \frac{{\sum\limits_{k = 1}^K {{{\log }_2}\left( {1 + \frac{{p_{j,k}^*{g_{j,k}}}}{{{N_0} + {I_{j,k}}}}} \right)} }}{{K{{\log }_2}\left( {1 + \frac{{N{\rm{ - }}B{\rm{ + 1}}}}{B}\frac{{{P_0}{G_j}}}{{{N_0}}}} \right) + \sum\limits_{k = 1}^K {{{\log }_2}\left( {1 + \frac{{p_{j,k}^*{g_{j,k}}}}{{{N_0} + {I_{j,k}}}}} \right)} }}
%\end{equation}
%\begin{equation}
%{R_t} \le \frac{{{{\log }_2}\left( {1 + \frac{{N{\rm{ - }}B{\rm{ + 1}}}}{B}\frac{{{P_0}{G_j}}}{{{N_0}}}} \right){{\log }_2}\left( {1 + \frac{{p_{j,k}^*{g_{j,k}}}}{{{N_0} + {I_{j,k}}}}} \right)}}{{K{{\log }_2}\left( {1 + \frac{{N{\rm{ - }}B{\rm{ + 1}}}}{B}\frac{{{P_0}{G_j}}}{{{N_0}}}} \right) + \sum\limits_{k = 1}^K {{{\log }_2}\left( {1 + \frac{{p_{j,k}^*{g_{j,k}}}}{{{N_0} + {I_{j,k}}}}} \right)} }}
%\end{equation}
\begin{equation}
{R_t} \le \min \left\{ {{\varphi _j}} \right\}
\end{equation}
where
\begin{equation}
{\varphi _j} = \frac{{{{\log }_2}\left( {1 + \frac{{N{\rm{ - }}B{\rm{ + 1}}}}{B}\frac{{{P_0}{G_j}}}{{{\sigma ^{\rm{2}}}}}} \right){{\log }_2}\left( {1 + \frac{{p_{j,k}^*{g_{j,k}}}}{{{\sigma ^{\rm{2}}} + {I_{j,k}}}}} \right)}}{{K{{\log }_2}\left( {1 + \frac{{N{\rm{ - }}B{\rm{ + 1}}}}{B}\frac{{{P_0}{G_j}}}{{{\sigma ^{\rm{2}}}}}} \right) + \sum\limits_{k = 1}^K {{{\log }_2}\left( {1 + \frac{{p_{j,k}^*{g_{j,k}}}}{{{\sigma ^{\rm{2}}} + {I_{j,k}}}}} \right)} }}.
\end{equation}

%Algorithm Design
\section{Algorithm Design}

According to the analysis of power allocation and wireless backhaul bandwidth allocation discussed above, we propose an iterative optimization algorithm and a suboptimal low-complexity algorithm.

\subsection{Iterative Resource Allocation Algorithm}

The proposed iterative resource allocation algorithm is shown in Algorithm 1.

\setcounter{algorithm}{0}
{\renewcommand\baselinestretch{1.0}\small
\begin{algorithm}[!h]\vspace{-1pt}
\caption{Iterative Resource Allocation Algorithm}
\begin{algorithmic}[1]{\small}
\STATE  \textbf{Initialization:} Each small cell BS allocates the same transmit power to each user, ${p_{j,k}} > 0$ and set $l = 1$.
\REPEAT
\STATE  \textbf{Backhaul Bandwidth Allocation}
\STATE  Compute optimum $\beta $ according to (\ref{computebeta}).
\STATE  Macro BS broadcasts the updated wireles backhaul bandwidth allocation factor to all small cell BSs.
\FOR    {each small cell BS}
\FOR    {each small cell user}
\STATE  \textbf{Power Allocation}
\STATE  a) find ${{\hat p}^*}_{j,k} = \arg \max {U_{j,k}}\left( {{p_{j,k}}} \right)$ according to GABS;
\STATE  b) check power constraint;
\IF     {${L_{j,k}} \le {{\hat p}^*}_{j,k} \le {H_{j,k}}$}
\STATE  $p_{j,k}^* = {{\hat p}^*}_{j,k}$
\ENDIF
\IF     {${{\hat p}^*}_{j,k} < {L_{j,k}}$}
\STATE  $p_{j,k}^* = {L_{j,k}}$
\ENDIF
\IF     {${{\hat p}^*}_{j,k} > {H_{j,k}}$}
\STATE  $p_{j,k}^* = {H_{j,k}}$
\ENDIF
\ENDFOR
\ENDFOR
\STATE  $l = l + 1$.
\UNTIL {total energy efficiency convergence or $l = {L_{\max }}$}
\end{algorithmic}
\end{algorithm}
\par}

In Algorithm 1, each small cell BS calculates ${\phi _j}$ according to \eqref{phi _j} and then sends ${\phi _j}$ to macro BS. The macro BS chooses the maximum ${\phi _j}$ to be the optimal bandwidth allocation factor $\beta $ and broadcasts $\beta $ to all small cell BSs.

\subsection{Low-Complexity Optimization Algorithm}

To reduce the complexity of Algorithm 1, we propose a low-complexity optimization algorithm where bandwidth allocation factor is calculated from the equal power allocation and we fix $\beta$ to calculate the power allocation according to the scheme proposed in Section IV. This low-complexity optimization algorithm is shown in Algorithm 2.

{\renewcommand\baselinestretch{1.0}\small
\begin{algorithm}[!h]\vspace{-1pt}
\caption{Fixed $\beta$ and Optimum Power Allocation Algorithm}
\begin{algorithmic}[1]{\small}
\STATE  \textbf{Initialization:} Each small cell BS allocates the same transmit power to each user, ${p_{j,k}} > 0$.
\STATE  \textbf{Backhaul Bandwidth Allocation}
\STATE  Compute optimum $\beta $ according to (\ref{computebeta}).
\STATE  Macro BS broadcasts the wireles backhaul bandwidth allocation factor to all small cell BSs.
\FOR    {each small cell BS}
\FOR    {each small cell user}
\STATE  \textbf{Power Allocation}
\STATE  a) find ${{\hat p}^*}_{j,k} = \arg \max {U_{j,k}}\left( {{p_{j,k}}} \right)$ according to GABS;
\STATE  b) check power constraint;
\IF     {${L_{j,k}} \le {{\hat p}^*}_{j,k} \le {H_{j,k}}$}
\STATE  $p_{j,k}^* = {{\hat p}^*}_{j,k}$
\ENDIF
\IF     {${{\hat p}^*}_{j,k} < {L_{j,k}}$}
\STATE  $p_{j,k}^* = {L_{j,k}}$
\ENDIF
\IF     {${{\hat p}^*}_{j,k} > {H_{j,k}}$}
\STATE  $p_{j,k}^* = {H_{j,k}}$
\ENDIF
\ENDFOR
\ENDFOR
\end{algorithmic}
\end{algorithm}
\par}

\subsection{Complexity Analysis}

Since the problem formulated in \eqref{originalobj} and \eqref{originalconstraints} is not convex, the only way to obtain the optimal solution is to use the method of exhaustion. If we assume that it costs $P$ operations to calculate ${r_{j,k}}$ and it costs $Q$ operations to calculate ${C_j}$, the complexity of checking $C2$ and $C3$ in \eqref{originalconstraints} entails $KP + K + Q$ operations and $P+1$ operations, respectively. If we assume that it costs $S$ operations to calculate ${U_{j,k}}$, the complexity of obtaining the total energy efficiency of all small cell users entails $JKS + \left( {J - 1} \right)\left( {K - 1} \right)$ operations. The total complexity of getting the value of objective function in \eqref{originalobj} under the constraints in \eqref{originalconstraints} entails $KP + K + P + 1 + JKS + \left( {J - 1} \right)\left( {K - 1} \right)$ operations for specific ${p_{j,k}}$ and $\beta $ values. If we assume the value step size for ${p_{j,k}}$ is $a$ and the value step size for $\beta $ is $b$, there are $\frac{1}{b}{\left( {\frac{{{P_{\max }}}}{a}} \right)^{JK}}$ choices for the values of ${p_{j,k}}$ and $\beta $. Therefore, the complexity for the method of exhaustion is $O\left( {\frac{{JKS}}{b}{{\left( {\frac{{{P_{\max }}}}{a}} \right)}^{JK}}} \right)$.

In Algorithm 1, the worst-case complexity of calculating bandwidth allocation factor $\beta$ from (\ref{computebeta}) entails $J$ operations in each iteration. If we assume that it costs $\Omega$ operations in each GABS to search the optimum power allocation without power constraint, then the worst-case complexity of finding the power allocation for every user in each small cell entails $JK\left( {\Omega {\rm{ + 4}}} \right)$ operations in each iteration. Suppose the Algorithm 1 needs $\Delta$ iterations to converge, so the total complexity of Algorithm 1 is $O\left( {JK\Omega \Delta } \right)$. Since iteration is not applied in Algorithm 2, the total complexity of Algorithm 2 is $O\left( {JK\Omega } \right)$, which is less than that of Algorithm 1. In the simulation, the typical value for $\Delta$ is around 16, the typical value for $\Omega $ is less than 500, and the typical values for $\frac{1}{b}$ and $\frac{{{P_{\max }}}}{b}$ are both 100. So the complexities of Algorithm 1 and Algorithm 2 are always less than that of the method of exhaustion. When the number of small cells $J$ and the number of users in each small cell $K$ increase, the complexity of the method of exhaustion increases exponentially, so the complexity of the method of exhaustion is much larger than the complexities of two proposed algorithms.

%Simulation Results
\section{Simulation Results}
Simulation results are given in this section to evaluate the performance of the proposed power allocation and backhaul bandwidth allocation algorithms. In the simulations, it is assumed that small cells are uniformly distributed in the macrocell coverage area, and small cell users are uniformly distributed in the coverage area of their serving small cell. AWGN power ${\sigma ^{\rm{2}}}{\rm{ = }}3.9811 \times {10^{ - 14}}$ W. The coverage radius of the macrocell is 500 m, and that of a small cell is 10 m. Small cell has a minimal distance of 50 m from the macro BS. The minimal distance between small cell BSs is 40 m. We assume that the channel fading is composed of path loss, shadowing fading, and Rayleigh fading. The pathloss model for small cell users is based on \cite{Further-Advancements10}. The lognormal shadowing between small cell BS and small cell users is 10 dB. At the macro BS, we assume that the transmit power is 33 dBm, the antenna array size $N=100$ and beamforming group size is $B=20$. We consider that all the small cell users have the same QoS requirement.

Figure 2 shows the convergence in terms of the energy efficiency of all small cell users for the proposed Algorithm 1 versus the number of iterations, where $J = 5$, ${R_t} = 0.01$ bps/Hz, ${P_{\max }} = 20$ dBm. It can be observed that the proposed resource allocation algorithm takes nearly 16 iterations to converge to stable solutions. This result, together with the previous analysis, ensures that the proposed Algorithm 1 is applicable in heterogeneous small cell networks.

Figure 3 shows the total energy efficiency of all small cell users when the number of users per small cell is increased from 2 to 10, for Algorithm 2 under ${P_{\max }} = 7$ dBm, ${P_{\max }} = 10$ dBm and ${P_{\max }} = 20$ dBm compared with Algorithm 1 under ${P_{\max }} = 20$ dBm. The simulation parameters are set as $J = 5$, ${R_t} = 0.01$ bps/Hz. Fig. 3 shows that the energy efficiency performance of Algorithm 1 is 20\% more superior to that of Algorithm 2. It also can be seen from Fig. 3 that the more number of users in small cell is, the better performance is obtained because of the multi-user diversity.

Figure 4 shows the total energy efficiency of all small cell users when the number of small cells is increased from 3 to 15, for Algorithm 2 under ${P_{\max }} = 7$ dBm, ${P_{\max }} = 10$ dBm and ${P_{\max }} = 20$ dBm when compared with Algorithm 1 under ${P_{\max }} = 20$ dBm. The simulation parameters are set as $K = 5$, ${R_t} = 0.01$ bps/Hz. Fig. 4 indicates that more number of small cell is, the better performance is obtained. It can also be seen from Fig. 4 that the energy efficiency performance of Algorithm 1 is always better than that of Algorithm 2 and the gap between them becomes larger when the number of small cells increases. The energy efficiency performance of Algorithm 1 is 30\% superior to that of Algorithm 2 when the number of small cells is 10.

Figure 5 shows the total downlink capacity of all small cell users when the number of users per small cell is increased from 2 to 10, for Algorithm 2 under ${P_{\max }} = 7$ dBm, ${P_{\max }} = 10$ dBm and ${P_{\max }} = 20$ dBm compared with Algorithm 1 under ${P_{\max }} = 20$ dBm. The simulation parameters are set as $J = 5$, ${R_t} = 0.01$ bps/Hz. Fig. 5 shows that the total downlink capacity of Algorithm 1 is more than 3 bps/Hz higher than that of Algorithm 2. It can also be seen from Fig. 5 that the more number of users in small cell is, the better performance is obtained due to the multi-user diversity. The total downlink capacity of Algorithm 1 is 21\% higher than that of Algorithm 2 when the number of users in each small cell is over 10.

Figure 6 shows the total downlink capacity of all small cell users when the number of small cells is increased from 3 to 15, for Algorithm 2 under ${P_{\max }} = 7$ dBm, ${P_{\max }} = 10$ dBm and ${P_{\max }} = 20$ dBm compared with Algorithm 1 under ${P_{\max }} = 20$ dBm. The simulation parameters are set as $K = 5$, ${R_t} = 0.01$ bps/Hz. Fig. 6 illustrates that Algorithm 1 is superior to Algorithm 2 in terms of the total downlink capacity and the gap between them becomes larger when the number of small cells increases. The total downlink capacity of Algorithm 1 is 29\% larger than that of Algorithm 2 when there are 14 small cells in the heterogeneous network.

Figure 7 shows the total energy efficiency of all the small cell users when using Algorithm 2 for power constraint ${P_{\max }}$ ranging from 0 dBm to 12.79 dBm where the number of users in each small cell is 3, 4, 5. The simulation parameters are set as $J = 5$, ${R_t} = 0.01$ bps/Hz. Fig. 7 presents that the more users in each small cell are, the higher total energy efficiency can be obtained, which has already been shown in Fig. 3. It can also be seen from Fig. 7 that the larger power constraint is, the better performance is obtained. This is because the larger power constraint leads to the larger region of the optimizing variable.

Figure 8 shows the total energy efficiency of all the small cell users when the number of users per small cell is increased from 2 to 10, for different algorithms. Algorithm 1 and Algorithm 2 are the iterative optimization algorithm and the low-complexity optimization algorithm, respectively, which we have proposed in Section IV. Algorithm 3 is an existing energy efficiency optimization algorithm with equal power allocation and Algorithm 4 is an algorithm that uses the optimum power allocation we proposed given a random $\beta$ to optimize energy efficiency. All the algorithms are under the constraint ${P_{\max }} = 20$ dBm. Fig. 8 indicates that more users in each small cell are, the better performance can be obtained, which has already been shown in Fig. 3. It also can be seen from Fig. 8 that Algorithm 1 has the best performance, and then it follows by Algorithm 2, Algorithm 3 and Algorithm 4. The energy efficiency performance of Algorithm 1 is 30.5\% and 56.6\% higher than that of Algorithm 3 and Algorithm 4, respectively.

Figure 9 shows the total energy efficiency of all the small cell users when the number of small cells is increased from 2 to 5, for the optimal solution and the two proposed algorithms. Since the complexity of the method of exhaustion is high, we only consider the situation with small dimension where there are two users located in each small cell, $K=2$. All the algorithms are under the setting of ${P_{\max }} = 20$ dBm and ${R_t} = 0.01$ bps/Hz. From Fig. 9, we can observe that the difference between the optimal solution and Algorithm 1 in terms of energy efficiency is small, which ensures the effectiveness of the proposed algorithms. The energy efficiency performance of the optimal solution is only about 7\% and 24\% higher than that of Algorithm 1 and Algorithm 2 respectively when the number of small cells is 3. The difference between the optimal solution and the proposed Algorithm 1 is mainly caused by the approximation of $C2$ in \eqref{P1.1constraints}. We can also observe that the performance of Algorithm 1 is slightly better than that of the existing algorithm, which is the backhaul bandwidth allocation in conjunction with the resource allocation algorithm in [14]. This phenomenon can be explained as follows. As the QoS requirement of small cell users increases, the more power is required to meet the higher QoS requirement.

%Conclusion
\section{Conclusion}

In this paper, we investigated the energy-efficient wireless backhaul bandwidth allocation and power allocation in a heterogeneous small cell network. We demonstrated the existence of a unique globally optimal energy efficiency solution and provided an iterative algorithm to obtain this optimum. For the downlink scenario, we first found the near optimal energy-efficient resource allocation approach, and then developed a low-complexity suboptimal algorithm by exploring the inherent structure of the objective function and the feature of energy-efficient design. From the simulation results, we observed that energy efficiency is improved by increasing the number of small cells and the number of users per small cell, and the capacity is also improved by increasing the number of small cells and the number of users per small cell. Simulation results showed great energy efficiency improvement of the proposed iterative optimization algorithm than that of the low-complexity optimization algorithm and the existing schemes. The proposed low-complexity algorithms can achieve a promising tradeoff between performance and complexity. If the future, we will investigate the nonunified backhaul bandwidth allocation and inter-small-cell interference in heterogeneous small cell networks.

\appendices

\section{Proof of Lemma 1}
We first focus on ${U_{j,k}}({p_{j,k}})$ and then we can obtain the properties of $U({\bf{P}})$. If every user in each small cell can reach the maximum energy efficiency, all small cell users can reach the maximum energy efficiency.

Denote the $\alpha$--superlevel sets of ${U_{j,k}}({p_{j,k}})$ as
\begin{equation}
{S_\alpha } = \{ \left. {{p_{j,k}} \ge 0} \right|{U_{j,k}}({p_{j,k}}) \ge \alpha \}
\end{equation}
where ${{p_{j,k}}}$ is nonnegative. Based on the propositions in \cite{Topics-in-Microeconomics99}, ${U_{j,k}}({p_{j,k}})$ is strictly quasiconcave if and only if ${S_\alpha }$ is strictly convex for any real number $\alpha $. In this case, when $\alpha  < 0$, no points exist on the contour ${U_{j,k}}({p_{j,k}}) = \alpha $. When $\alpha  = 0$, only ${p_{j,k}} = 0$ is on the contour ${U_{j,k}}(0) = \alpha$. Hence, ${S_\alpha }$ is strictly convex when $\alpha  \le 0$. Now, we investigate the case when $\alpha  > 0$. We can rewrite the ${S_\alpha }$ as ${S_\alpha } = \{ \left. {{p_{j,k}} \ge 0} \right|\alpha {P_C} + \alpha {p_{j,k}} - {r_{j,k}}({p_{j,k}}) \le 0\}$. Since ${r_{j,k}}({p_{j,k}})$ is strictly concave in ${{p_{j,k}}}$, $ - {r_{j,k}}({p_{j,k}})$ is strictly convex in ${{p_{j,k}}}$; therefore, ${S_\alpha }$ is strictly convex. Hence, we have the strict quasiconcavity of ${U_{j,k}}({p_{j,k}})$.

Next, we can obtain the partial derivative of ${U_{j,k}}({p_{j,k}})$ with ${p_{j,k}}$ as
\begin{equation}
\frac{{\partial {U_{j,k}}({p_{j,k}})}}{{\partial {p_{j,k}}}} = \frac{{({P_C} + {p_{j,k}}){{r'}_{j,k}}({p_{j,k}}) - {r_{j,k}}({p_{j,k}})}}{{{{\left( {{P_C} + {p_{j,k}}} \right)}^2}}}{\rm{ = }}\frac{{f({p_{j,k}})}}{{{{\left( {{P_C} + {p_{j,k}}} \right)}^2}}}
\end{equation}
where $f({p_{j,k}}) = ({P_C} + {p_{j,k}}){{r'}_{j,k}}({p_{j,k}}) - {r_{j,k}}({p_{j,k}})$, ${{r'}_{j,k}}({p_{j,k}})$ is the first partial derivative of ${r_{j,k}}({p_{j,k}})$ with respect to ${p_{j,k}}$. If $p_{j,k}^*$ exists such that ${\left. {\frac{{\partial {U_{j,k}}({p_{j,k}})}}{{\partial {p_{j,k}}}}} \right|_{{p_{j,k}} = p_{j,k}^*}} = 0$, it is unique, i.e. if there is a $p_{j,k}^*$ such that $f(p_{j,k}^*) = 0$. In the following, we investigate the conditions when $p_{j,k}^*$ exists.

The derivative of $f({p_{j,k}})$ is
\begin{equation}
f'({p_{j,k}}) = ({P_C} + {p_{j,k}}){{r''}_{j,k}}({p_{j,k}})
\end{equation}
where ${{r''}_{j,k}}({p_{j,k}})$ is the second partial derivative of ${r_{j,k}}({p_{j,k}})$ with respect to ${p_{j,k}}$. Since ${r_{j,k}}({p_{j,k}})$ is strictly concave in ${{p_{j,k}}}$, so ${{r''}_{j,k}}({p_{j,k}}) < 0$, $f'({p_{j,k}}) < 0$. Hence, $f({p_{j,k}})$ is strictly decreasing.
\begin{equation}
  \begin{split}
    \mathop {\lim }\limits_{{p_{j,k}} \to \infty } f({p_{j,k}})&=\mathop {\lim }\limits_{{p_{j,k}} \to \infty } \left( {({P_C} + {p_{j,k}}){{r'}_{j,k}}({p_{j,k}}) - {r_{j,k}}({p_{j,k}})} \right)\\
    &=\mathop {\lim }\limits_{{p_{j,k}} \to \infty } \left( {{P_C}{{r'}_{j,k}}({p_{j,k}}) + {p_{j,k}}{{r'}_{j,k}}({p_{j,k}}) - {r_{j,k}}({p_{j,k}})} \right)
  \end{split}
\end{equation}
where
\begin{equation}
{{r'}_{j,k}}({p_{j,k}}) = \left( {\frac{{1 - \beta }}{K}} \right)\left( {\frac{{{g_{j,k}}}}{{{\sigma ^{\rm{2}}} + {I_{j,k}}}}} \right)\left( {\frac{1}{{\ln 2}}} \right)\left( {\frac{1}{{1 + \frac{{{p_{j,k}}{g_{j,k}}}}{{{\sigma ^{\rm{2}}}{\rm{ + }}{I_{j,k}}}}}}} \right)
\end{equation}
and
\begin{equation}
\mathop {\lim }\limits_{{p_{j,k}} \to \infty } {{r'}_{j,k}}({p_{j,k}}) = 0.
\end{equation}
So we have
\begin{equation}
\mathop {\lim }\limits_{{p_{j,k}} \to \infty } {P_C}{{r'}_{j,k}}({p_{j,k}}) = 0.
\end{equation}
According to the L'Hopital's rule, it is easy to show that
\begin{equation}
  \begin{split}
    \mathop {\lim }\limits_{{p_{j,k}} \to \infty } {p_{j,k}}{{r'}_{j,k}}({p_{j,k}})& = \mathop {\lim }\limits_{{p_{j,k}} \to \infty } \left( {\frac{{1 - \beta }}{K}} \right)\left( {\frac{{{g_{j,k}}}}{{{\sigma ^{\rm{2}}} + {I_{j,k}}}}} \right)\left( {\frac{1}{{\ln 2}}} \right)\left( {\frac{{{p_{j,k}}}}{{1 + \frac{{{p_{j,k}}{g_{j,k}}}}{{{\sigma ^{\rm{2}}}{\rm{ + }}{I_{j,k}}}}}}} \right)\\
    & = \mathop {\lim }\limits_{{p_{j,k}} \to \infty } \left( {\frac{{1 - \beta }}{K}} \right)\left( {\frac{{{g_{j,k}}}}{{{\sigma ^{\rm{2}}} + {I_{j,k}}}}} \right)\left( {\frac{1}{{\ln 2}}} \right)\left( {\frac{1}{{\frac{{{g_{j,k}}}}{{{\sigma ^{\rm{2}}} + {I_{j,k}}}}}}} \right)\\
    &=\mathop {\lim }\limits_{{p_{j,k}} \to \infty } \left( {\frac{{1 - \beta }}{K}} \right)\left( {\frac{1}{{\ln 2}}} \right)
  \end{split}
\end{equation}
\begin{equation}
\mathop {\lim }\limits_{{p_{j,k}} \to \infty } \left( { - {r_{j,k}}({p_{j,k}})} \right) = \mathop {\lim }\limits_{{p_{j,k}} \to \infty } \left[ { - \left( {\frac{{1 - \beta }}{K}} \right){{\log }_2}\left( {1 + \frac{{{p_{j,k}}{g_{j,k}}}}{{{\sigma ^{\rm{2}}}{\rm{ + }}{I_{j,k}}}}} \right)} \right] =  - \infty.
\end{equation}
So we have
\begin{equation}
\mathop {\lim }\limits_{{p_{j,k}} \to \infty } f({p_{j,k}}) < 0.
\end{equation}
Besides,
\begin{equation}
  \begin{split}
    \mathop {\lim }\limits_{{p_{j,k}} \to 0} f({p_{j,k}})&=\mathop {\lim }\limits_{{p_{j,k}} \to 0} \left( {({P_C} + {p_{j,k}}){{r'}_{j,k}}({p_{j,k}}) - {r_{j,k}}({p_{j,k}})} \right)\\
    &= {P_C}{{r'}_{j,k}}(p_{j,k}^{(0)}) - {r_{j,k}}(p_{j,k}^{(0)})\\
  \end{split}
\end{equation}
where $p_{j,k}^{(0)}$ denotes ${p_{j,k}} = 0$

%\begin{equation}
%{\bf{P}}_{j,k}^{(0)}=
%\begin{bmatrix}
%     {{p_{1,1}}} & \cdots & {{p_{1,k - 1}}} & {{p_{1,k}}}& {{p_{1,k + 1}}} & \cdots& {{p_{1,K}}}\\
%     \vdots & \ddots & \vdots & \ddots & \vdots & \ddots & \vdots \\
%     {{p_{j,1}}} & \cdots & {{p_{j,k - 1}}} & 0 & {{p_{j,k + 1}}} & \cdots & {{p_{j,K}}}\\
%     \vdots & \ddots & \vdots & \ddots & \vdots & \ddots & \vdots \\
%     {{p_{J,1}}} & \cdots & {{p_{J,k - 1}}} & {{p_{J,k}}}& {{p_{J,k + 1}}} & \cdots & {{p_{J,K}}}\\
%\end{bmatrix}
%\end{equation}

\begin{equation}
  \begin{split}
    {{r'}_{j,k}}(p_{j,k}^{(0)})& = {\left. {\left( {\frac{{1 - \beta }}{K}} \right)\left( {\frac{{{g_{j,k}}}}{{{\sigma ^{\rm{2}}} + {I_{j,k}}}}} \right)\left( {\frac{1}{{\ln 2}}} \right)\left( {\frac{1}{{1 + \frac{{{p_{j,k}}{g_{j,k}}}}{{{\sigma ^{\rm{2}}}{\rm{ + }}{I_{j,k}}}}}}} \right)} \right|_{{p_{j,k}} = 0}}\\
    & = \left( {\frac{{1 - \beta }}{K}} \right)\left( {\frac{{{g_{j,k}}}}{{{\sigma ^{\rm{2}}} + {I_{j,k}}}}} \right)\left( {\frac{1}{{\ln 2}}} \right)
  \end{split}
\end{equation}
\begin{equation}
{r_{j,k}}(p_{j,k}^{(0)}) = 0
\end{equation}
\begin{equation}
\mathop {\lim }\limits_{{p_{j,k}} \to 0} f({p_{j,k}}) = \left( {\frac{{1 - \beta }}{K}} \right)\left( {\frac{{{P_C}{g_{j,k}}}}{{{\sigma ^{\rm{2}}} + {I_{j,k}}}}} \right)\left( {\frac{1}{{\ln 2}}} \right) > 0.
\end{equation}

So, together with $\mathop {\lim }\limits_{{p_{j,k}} \to \infty } f({p_{j,k}}) < 0$, we see that ${p^ * }_{j,k}$ exists and ${U_{j,k}}({p_{j,k}})$ is first strictly increasing and then strictly decreasing in ${p_{j,k}}$.

Lemma 1 is readily obtained.
$\hfill\blacksquare$

\newpage
\begin{figure}[h]
        \centering
        \includegraphics*[width=13cm]{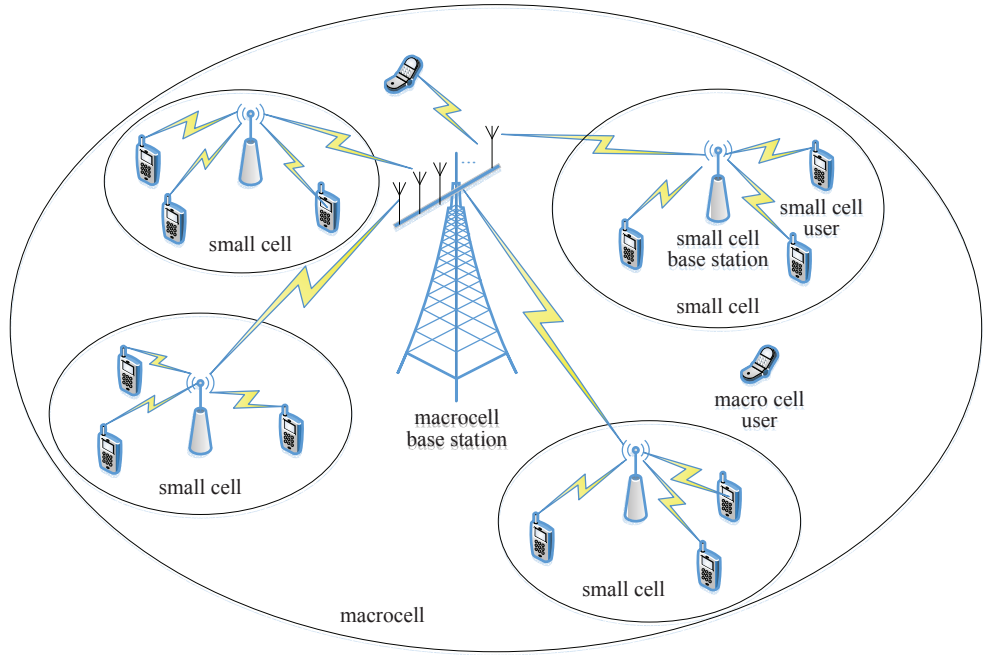}
        \caption{Topology of a heterogeneous small cell network.}
        \label{fig:1}

\end{figure}

\begin{figure}[h]
        \centering
        \includegraphics*[width=13cm]{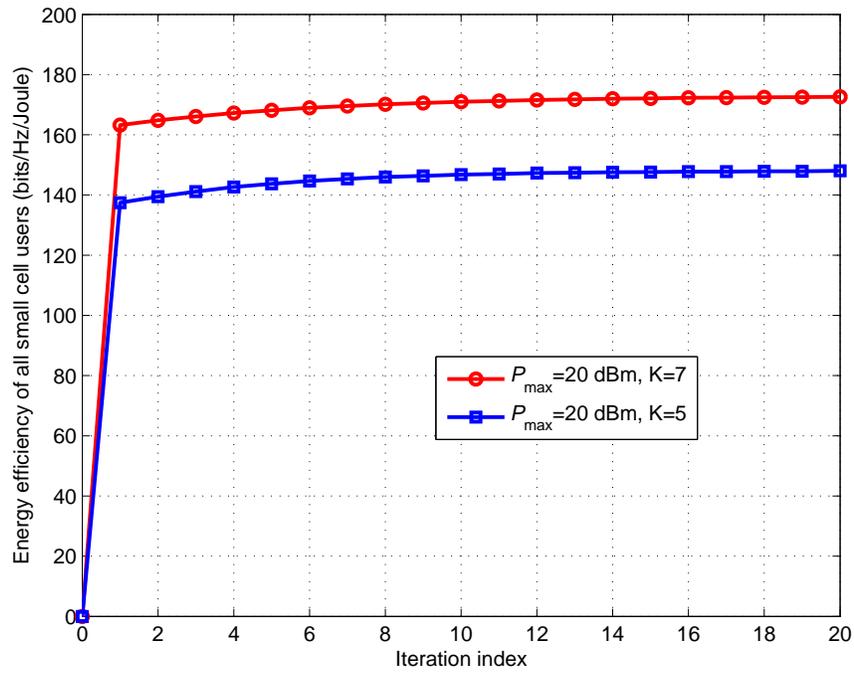}
        \caption{The convergence in terms of energy efficiency of all small cell users over the number of iterations.}
        \label{fig:2}

\end{figure}

\begin{figure}[h]
        \centering
        \includegraphics*[width=13cm]{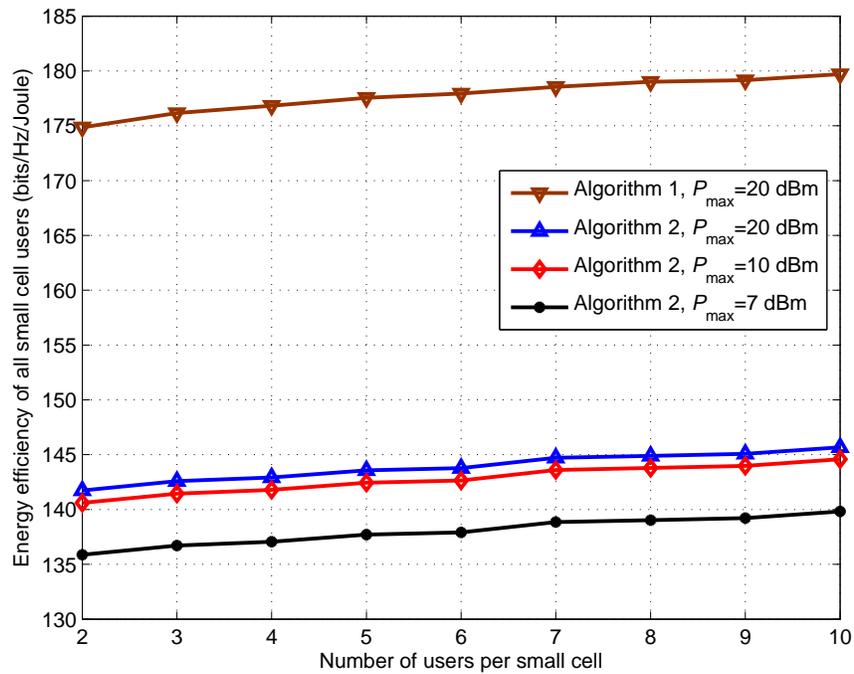}
        \caption{Energy efficiency versus the number of users per small cell.}
        \label{fig:3}

\end{figure}

\begin{figure}[h]
        \centering
        \includegraphics*[width=13cm]{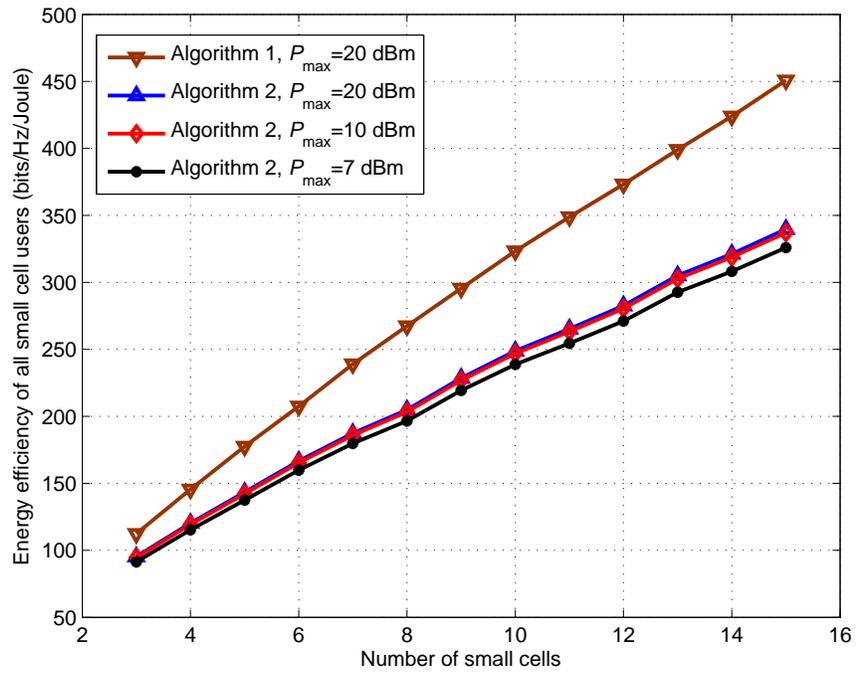}
        \caption{Energy efficiency versus the number of small cells.}
        \label{fig:4}

\end{figure}

\begin{figure}[h]
        \centering
        \includegraphics*[width=13cm]{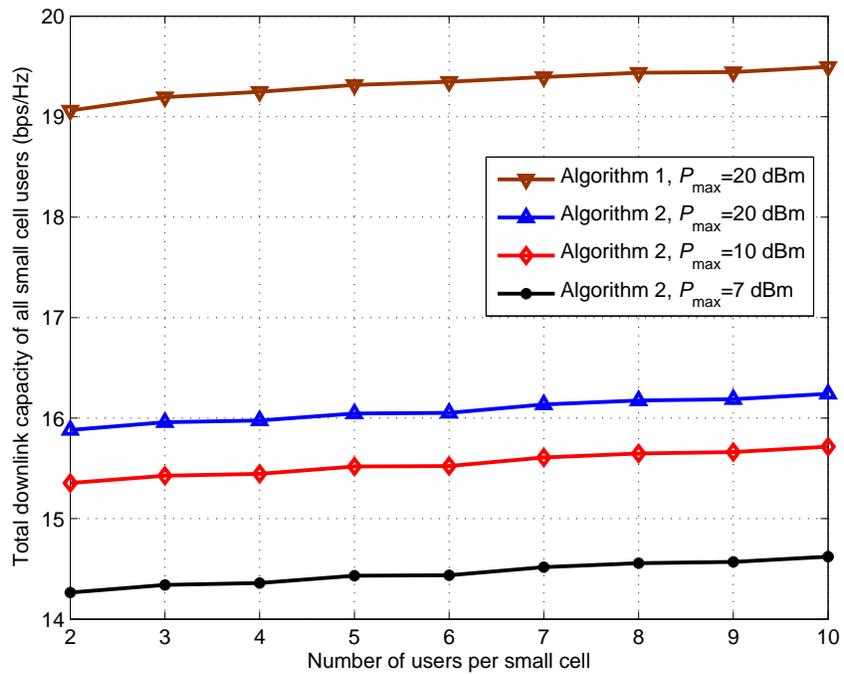}
        \caption{Capacity versus the number of users per small cell.}
        \label{fig:5}

\end{figure}

\begin{figure}[h]
        \centering
        \includegraphics*[width=13cm]{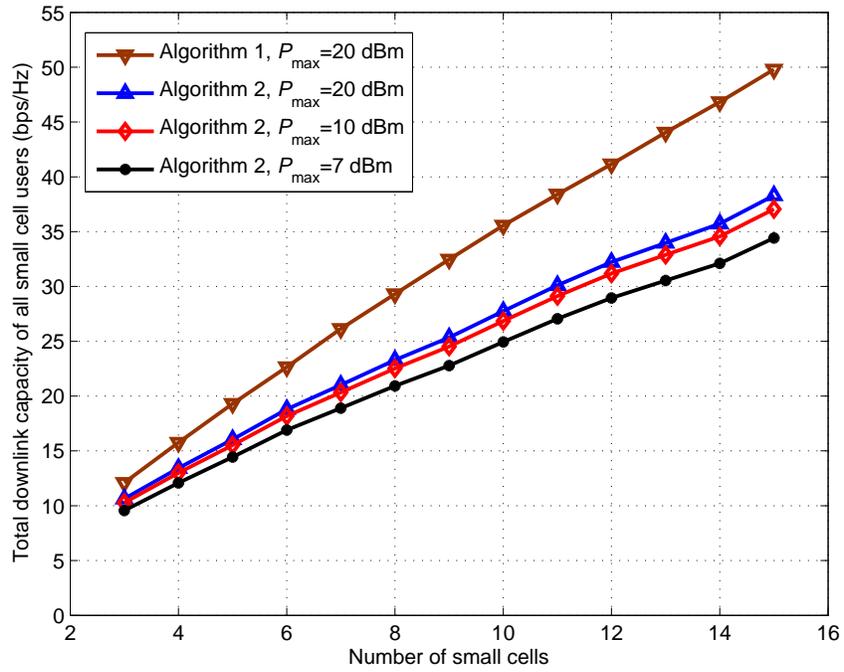}
        \caption{Capacity versus the number of small cells.}
        \label{fig:6}

\end{figure}

\begin{figure}[h]
        \centering
        \includegraphics*[width=13cm]{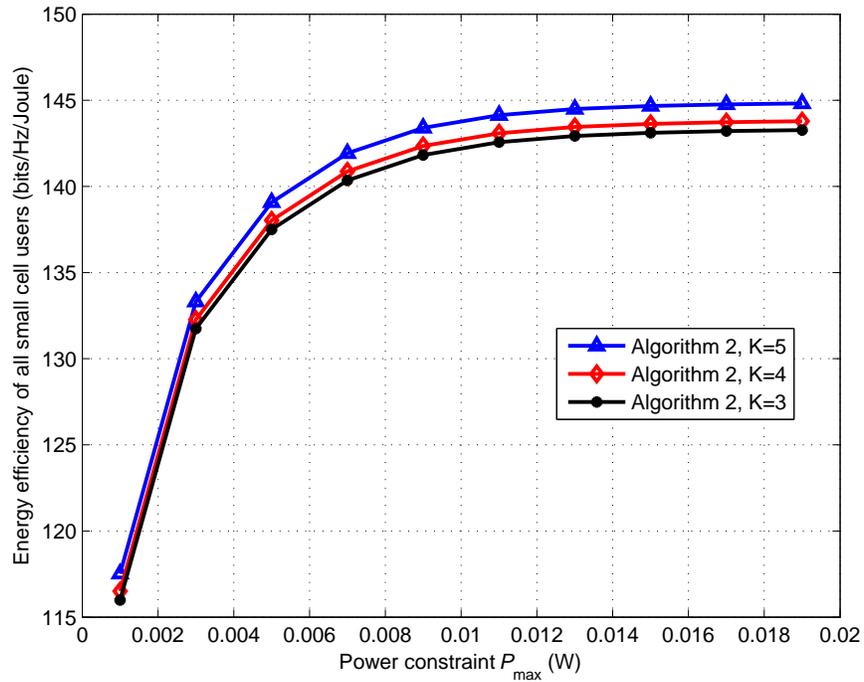}
        \caption{Energy efficiency versus the power constraint.}
        \label{fig:7}

\end{figure}

\begin{figure}[h]
        \centering
        \includegraphics*[width=13cm]{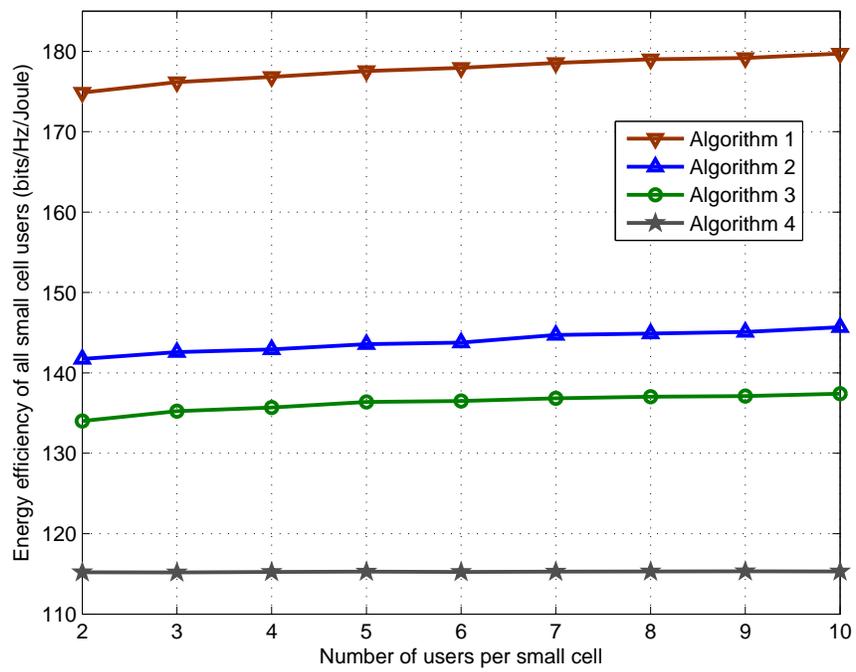}
        \caption{Energy efficiency comparison for different algorithms.}
        \label{fig:8}

\end{figure}

\begin{figure}[h]
        \centering
        \includegraphics*[width=13cm]{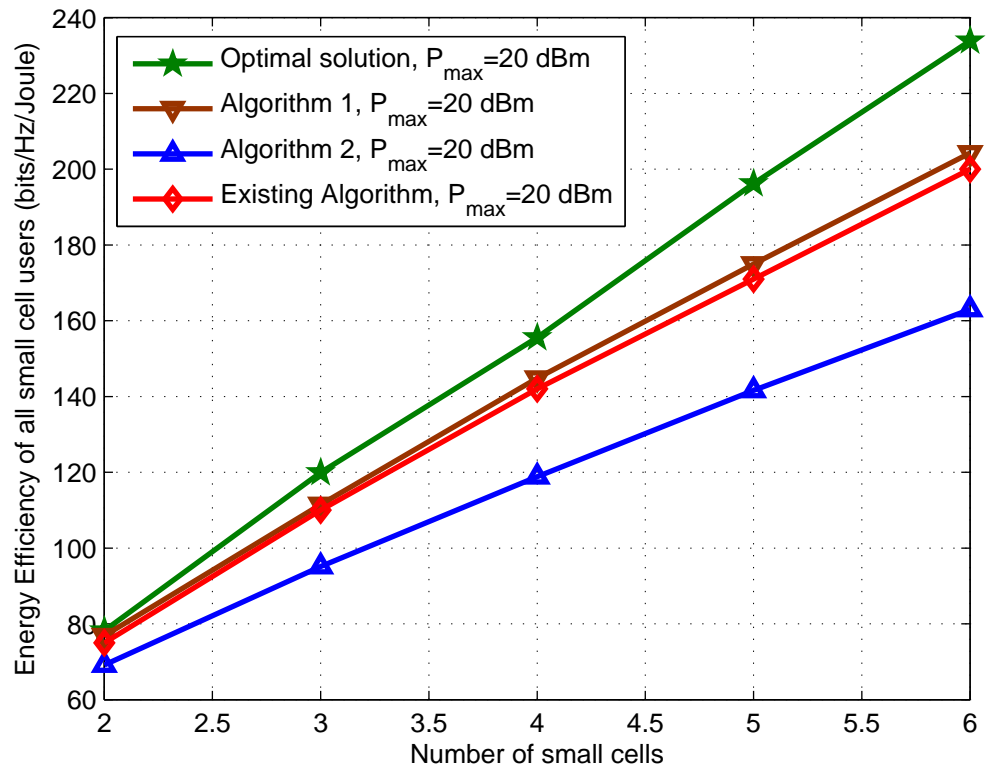}
        \caption{Energy efficiency comparison for the optimal solution and proposed algorithms.}
        \label{fig:9}

\end{figure}

\end{document}